# Ammonium thiocyanate enables *in-situ* structural templating of $MAPbI_3$ *via* quasi-2D perovskite scaffold

Sophie Tucker[1,2], Isaac Metcalf[3,4], Rachel C. Kilbride[5,6], Elena J. Cassella[1], Jessica H. Persaud[3], Faiz Mandani[4], Niranjana Raj[7], Max Allen[1,2], Charlie H. G. Nicholls[8], David G. Lidzey[1], Alice L. B. Pyne[8], Robert D. J. Oliver[8], Silvia Motti[7], Jacky Even[9], Aditya D. Mohite[3,4], Alexandra J. Ramadan[1]*.

[1] School of Mathematical and Physical Sciences, University of Sheffield, Hicks Building, Hounsfield Road, Sheffield S3 7RH, UK
[2] Grantham Centre for Sustainable Futures, University of Sheffield, Sheffield S10 2TN, UK
[3] Department of Chemical and Biomolecular Engineering, Rice University, Houston, TX 77005, USA
[4] Department of Materials Science and NanoEngineering, Rice University, Houston, TX 77005, USA
[5] XMaS: The UK X-ray Materials Science Facility at the European Synchrotron Radiation Facility, Grenoble F-38043, France
[6] Department of Physics, University of Warwick, Coventry CV4 7AL, United Kingdom
[7] School of Physics and Astronomy, Faculty of Engineering and Physical Sciences, University of Southampton, University Road, Southampton SO17 1BJ, UK
[8] School of Chemical, Materials and Biological Engineering, University of Sheffield, Sir Robert Hadfield Building, Mappin Street, Sheffield, S1 3JF, UK
[9] Institut National des Sciences Appliquées (INSA) Rennes, Université Rennes, CNRS, Institut Fonctions Optiques pour les Technologies de l'Information (FOTON)–UMR 6082, F-35000 Rennes, France

*corresponding author: a.ramadan@sheffield.ac.uk

## Abstract:

Ammonium thiocyanate ($NH_4SCN$) has been used extensively to improve the current extraction and device performance of quasi-two-dimensional perovskite solar cells. Whilst this additive has been shown to influence quasi-2D perovskite crystallisation, a detailed understanding of its mechanism has not yet been established. Through spectroscopic and structural studies, we demonstrate that $NH_4SCN$ enables quasi-2D perovskite phases to structurally template the dominant 3D phase. Contrary to previous reports, we suggest that the benefit of this additive comes from an improvement to the 3D perovskite structure, not from charges transferring through vertically oriented 2D perovskite layers. We show that this additive provides kinetic control over the quasi-2D phase formation, suppressing the unfavourable $n$ = 1 perovskite phase and promoting the growth of high-$n$ 2D scaffold phases by extending the nucleation window. This promotes the *in-situ* templated growth of $MAPbI_3$. We provide robust structural analysis demonstrating highly (*110*)-ordered 3D perovskite, with the mechanism confirmed by close lattice matching between this phase and quasi-2D phases. Furthermore, we show that this structural control is cation dependent, requiring both close lattice matching between quasi-2D and 3D perovskite phases, and successful kinetic control of the relevant quasi-2D phases.

## Introduction:

Quasi-two-dimensional (quasi-2D) Ruddlesden-Popper perovskites are an emerging class of perovskites with improved stability in the presence of moisture[1–4], owing to the intercalation of bulky hydrophobic organic molecules along the (*100*) direction of the perovskite lattice. Ruddlesden-Popper perovskites have the general formula $R_2A_{n-1}B_nX_{3n+1}$, with $R$ being the bulky organic spacer cation (typically phenethylammonium or *n*-butylammonium), $A$ being the small perovskite 'cage' cation (methylammonium, formamidinium, or $Cs^+$), $B$ the metal cation ($Pb^{2+}$) and $X$ the halide ($I^-$). $n$ refers to the 'layer number', or the number of Pb atomic sheets confined between organic layers. Specifically, 'quasi-2D' refers to materials with $n > 1$, whereby a number of specific 2D phases are present. In these cases, the average layer number, $<n>$, can be used to reflect the stoichiometry of the overall material.

The cage cation in traditional three-dimensional (3D) perovskites is restricted by the Goldschmidt tolerance factor[5] meaning that there are only a small number of cations able to form a perovskite structure. Conversely, the spacer cation in quasi-2D perovskites is less limited by size as it does not sit within the perovskite octahedra, therefore lifting the restrictions on molecular structure. This awards quasi-2D perovskites a high degree of structural diversity[6,7], allowing control over properties that arise from spacer cation interactions[8]. In addition to their improved stability, quasi-2D perovskites share many of the optoelectronic properties of 3D perovskites that make them excellent candidates for photovoltaics (PV)[9,10], light emitting diodes[11,12], and field effect transistors[13], such as high absorption coefficients[14] and defect tolerance[15]. In the case of perovskite PV, improved moisture stability is important for the industrial relevance of these technologies in operational conditions[16].

Despite this stability benefit, the power conversion efficiency of quasi-2D perovskite PV lags behind their 3D counterparts. This has been ascribed to a reduction in out-of-plane charge mobility caused by the quasi-2D perovskite sheets preferentially orienting parallel to the substrate[17,18], and the organic cation layer acting as a barrier to charge transport.[19,20] In addition, the exciton binding energies in quasi-2D perovskites are increased by an order of magnitude (>100 meV[21] vs ~10 meV[22] for 3D perovskites) due to an increase in both quantum and dielectric confinement.[8,23] This results in an increased activation barrier to exciton dissociation, further reducing the charge extraction and therefore short-circuit current ($J_{SC}$). The confinement effects induced by the bulky cation layer increase the bandgap of these materials,[24] leading to reduced light absorption in the near-infrared and further inhibiting $J_{SC}$. These effects mean that for quasi-2D perovskite solar cells with $n < 5$, power conversion efficiencies have yet to compete with those of 3D perovskites, with record efficiencies currently held at ~21%[25]. Despite the reduced out-of-plane charge mobility, quasi-2D perovskites exhibit excellent in-plane charge mobility[26], prompting efforts to attempt to induce vertical growth of 2D sheets such that the perovskite layers form normal to the substrate[4]. Additionally, edge states were demonstrated to efficiently act as exciton dissociation centres for 2D sheets with vertical orientation embedded in a solar cell architecture[27]. This thereby allows charges to transport along, rather than across, the stacking layer, mitigating $J_{SC}$ losses. Thus far, avenues to induce vertical orientation have included kinetically controlling crystallisation *via* hot-casting[28] or vacuum deposition[29], solvent engineering[30], spacer cation tuning[31], and the use of additives[32,33].

Ammonium thiocyanate ($NH_4SCN$) is commonly used as an additive in both 3D and quasi-2D perovskites. In 3D perovskites, it is thought to improve perovskite formation through increasing grain size by delaying perovskite nucleation, and passivating defects by the inclusion of $SCN^-$ ions into $I^-$ vacancies in the perovskite lattice[34,35]. In quasi-2D perovskites, $NH_4SCN$ is typically used in the synthesis of structures featuring aromatic spacer cations, such as p-xylylene diamine ($XDA^+$)[36], phenethylammonium ($PEA^+$)[37], and 4-fluoro-phenethylammonium ($FPEA^+$)[38]. The use of $NH_4SCN$ increases charge extraction and boosts the $J_{SC}$ of quasi-2D perovskite solar cells. This is largely attributed to the additive inducing vertical orientation[39] in the target 2D phase, though the mechanism of this is not well understood.

Additionally, some reports[40] have suggested that the inclusion of $NH_4SCN$ into the quasi-2D perovskite precursor solution influences the specific structures that comprise the final film. While a quasi-2D perovskite precursor may be stoichiometrically prepared to reflect a particular $n$ structure (or 'phase'), the resultant film is often a combination of domains corresponding to different phases. For example, a solution prepared as $n = 4$ may result in a film containing regions of perovskite structured with $n = 1$, 2, 3, and 4, as well as domains resembling the 3D perovskite structure. This inhomogeneity can be detrimental to device performance and stability[41]. As the bandgap of the material varies with $n$, this can lead to charge funnelling effects caused by charges preferentially moving to regions with narrower bandgap (higher $n$ number)[41]. The presence of low-$n$ domains can act as sites for charge recombination and inhibit charge transport, acting as charge blocking layers[42,43]. In addition, the increase in the density of domain boundaries (between regions of differing $n$ number) can result in an increase in defect density, with defects being introduced in regions with lattice mismatch[44]. Therefore, controlling phase purity is crucial to the development of quasi-2D perovskite photovoltaics.

Understanding the role of ammonium thiocyanate in controlling both orientation and phase distribution is necessary to elucidate a better understanding of current extraction in quasi-2D perovskite photovoltaics. In this work, we show that the introduction of ammonium thiocyanate into the quasi-2D perovskite precursor solution controls the film crystallisation and subsequent phase distribution in <*n*> = 4 Ruddlesden-Popper perovskites. Contrary to previous reports, we posit that quasi-2D perovskites treated with $NH_4SCN$ favour the formation of the 3D perovskite phase, with the additive acting to inhibit the bulk formation of low-*n* 2D phases. We further show that the additive enables a templated growth of 3D perovskite, resulting in a more crystalline, ordered material and therefore higher quality perovskite. Additionally, we investigate the role of the spacer cation on the effect of $NH_4SCN$, and we determine that the additive-spacer interaction is critical in the control of phase distribution in quasi-2D perovskites. This work not only elucidates crucial information about the effect of $NH_4SCN$ on quasi-2D perovskites, but also sets the groundwork for the rational design of additives for a variety of 2D perovskite structures.

## Results and discussion:

To examine the role of $NH_4SCN$ in the formation of quasi-2D perovskites, we fabricated Ruddlesden-Popper perovskites using the aromatic spacer cations 4-fluoro-phenylammonium (FPA, $FC_6H_4NH_3^+$), 4-fluoro-benzylammonium (FBZA, $FC_6H_4CH_2NH_3^+$), and 4-fluoro-phenethylammonium (FPEA, $FC_6H_4CH_2CH_2NH_3^+$). The structures of these molecules are shown in **Figure S1a** and **Figure 1a**, respectively. These cations possess similar chemical structures but with differences in the length of the alkyl chain. Precursor solutions were prepared in dimethylformamide (DMF) according to the stoichiometric ratio for $n$ = 4, i.e., $R_2MA_3Pb_4I_{13}$ with $R$ being the spacer cation, MA = methylammonium ($CH_3NH_3^+$). The detailed masses are shown in the methods. A range of $NH_4SCN$ concentrations were screened, with 25 mol% with respect to Pb found to yield the highest photovoltaic performance (**Figure S2**), with higher concentrations leading to devices with electrical shorts due to incomplete film growth. Hence, this concentration was selected for all further studies, referred to hereafter as '$NH_4SCN$'. Thin films were fabricated *via* a one-step spin-coating process on quartz glass substrates (see **Experimental Methods** for more information).

Contrary to FBZA and FPEA, we found that samples prepared using FPA did not form a 2D or quasi-2D perovskite structure, but rather crystallised as $MAPbI_3$. This was evident in the UV-Visible absorption spectra, shown in **Figure S1b**, which displays evidence of an absorption onset at ~750 nm and no visible peaks associated with the Ruddlesden-Popper phases. Additionally, the photoluminescence (PL) spectra (**Figure S1c**) showed emission exclusively from the 3D phase at 780 nm. We suggest that this is due to steric hindrance, with the smaller cation being less able to interrupt the Ostwald ripening process from 2D to 3D perovskite phases, thereby rendering the 2D phases thermodynamically inaccessible[45]. Hence, we proceeded to investigate only the FPEA and FBZA perovskites.

We carried out in-situ PL measurements to observe the formation of the perovskites during spin-coating, with and without the influence of $NH_4SCN$. The PL emission was measured from the time that the spin-coating commenced. The maximum emission intensity over time is shown in **Figure 1b-c.** The onset of intensity is assumed to be the point at which the solution becomes supersaturated and the perovskite begins to crystallise[46]. The time for the emission to reach half the maximum intensity is marked with a dashed line. In both the FPEA and FBZA systems, the addition of the additive delays the emission, by 8.5 s for FPEA and 3.5 s for FBZA. This is accompanied by a corresponding delay in the colour change observed during spin-coating, in which the material changes from yellow to dark red. This suggests that the additive delays crystallisation, potentially due to the formation of intermediates that prevent nucleation from occurring, as has been shown previously with thiocyanates in 3D perovskites[47–49]. This is further corroborated by the fact that ammonium thiocyanate concentrations in excess of 50 mol% resulted in incomplete film growth and shorted devices (**Figure S2**). The delay is more significant in the FPEA perovskite compared with FBZA, suggesting that the spacer cation may play an active role in this interaction.

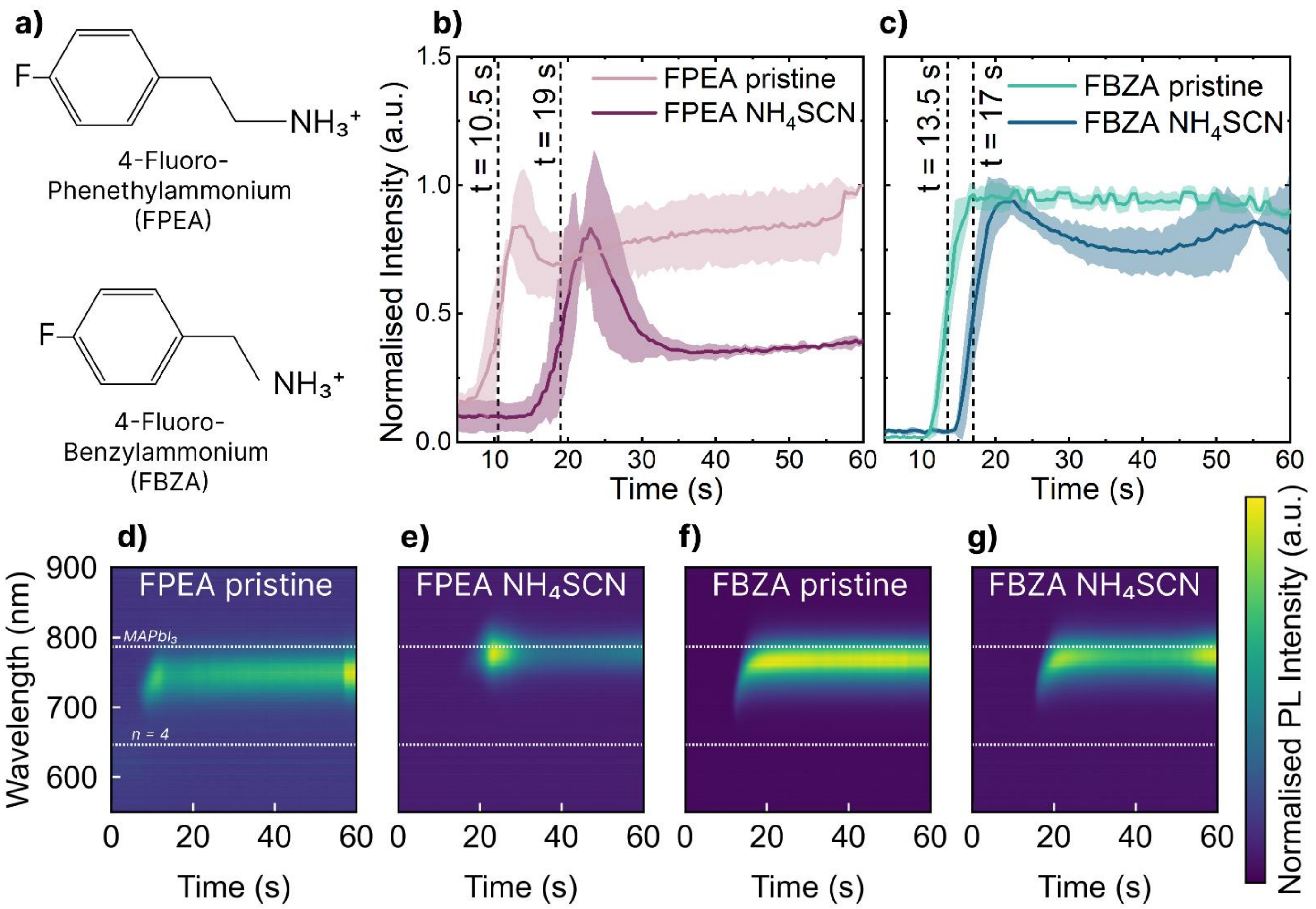

**Figure 1:** a) Molecular structures of 4-Fluoro-Phenethylammonium (FPEA) and 4-Fluoro-Benzylammonium (FBZA). b-c) Maximum photoluminescence emission intensity over time during spin-coating for b) $FPEA_2MA_3Pb_4I_{13}$ films with (dark pink) and without (light pink) $NH_4SCN$ and c) $FBZA_2MA_3Pb_4I_{13}$ films with (dark blue) and without (light blue) $NH_4SCN$. The solid line represents the average of three samples, and the shaded area represents the standard deviation. The dashed line indicates half the maximum intensity. d-g) Wavelength resolved, normalised photoluminescence emission spectra over time during spin-coating for d) pristine FPEA, e) $NH_4SCN$ FPEA, f) pristine FBZA, and g) $NH_4SCN$ FBZA films. Data is normalised such that all samples have the same maximum emission. White dashed lines show the expected emission of pure $MAPbI_3$ (top, 785 nm) and pure $n$ = 4 2D perovskite (bottom, 645 nm).

**Figure 1d-g** show maps of the intensity of emission over time with respect to wavelength. Notably, in all four perovskite systems the emission is centred closer to the emission of 3D $MAPbI_3$ than that of $n$ = 4 perovskite, around 750 - 770 nm. This is indicative of emission dominated by carriers recombining radiatively in 3D domains within the film. This suggests that if 2D or quasi-2D domains are present within the film they are not emissive, either due to structural defects preventing radiative recombination or due to a rapid funnelling of charges into lower energy 3D domains. At the onset of emission, all samples demonstrate initial photoluminescence that is slightly blue-shifted before stabilising at the final emission wavelength. This is coupled with a drop in PL intensity after early stages of growth. These features could suggest initial formation of perovskite nanocrystals as an intermediate stage of growth[50–52]. The addition of $NH_4SCN$ additionally induces a redshift of emission in both perovskite structures.

To investigate the effect of $NH_4SCN$ on post-annealed perovskite films, we measured the optical properties of the same perovskites on quartz coated glass. UV-visible absorption spectra of each condition are shown in **Figure 2a-b**. The pristine FPEA (not containing $NH_4SCN$) exhibits excitonic peaks associated with 2D phases $n$ = 1 - 4 at 520, 570, 610, and 650 nm respectively, as well as a 3D absorption onset at 765 nm. The FPEA sample with $NH_4SCN$ condition does not display an excitonic feature at 520 nm, suggesting that the $n$ = 1 phase is inhibited. The intensity of the excitonic peaks associated with the $n$ = 2 - 4 phases are increased compared to the pristine sample, suggesting that these intermediate phases are enhanced. Additionally, the 3D absorption onset at 765 nm is more prominent, potentially indicative of an increase of the crystallinity or domain size of the 3D perovskite phase in the film.

In the case of the FBZA perovskite, no such reduction in the $n$ = 1 phase is observed, although the $n$ = 2 phase is similarly enhanced, and the 3D onset is also larger with the addition of $NH_4SCN$. The persistence of FBZA $n$ = 1, despite a slower crystallization rate which should bring the film's phase distribution closer to equilibrium, suggests that the FBZA $n$ = 1 phase may be thermodynamically favoured over high-$n$-value phases. The peaks associated with the $n$ = 3 - 4 phases are absent from both pristine and $NH_4SCN$-treated samples, suggesting that these phases may not be thermodynamically stable. This suggests that the additive is less effective in changing the formation dynamics of the FBZA perovskite, highlighting the role of the spacer cation in this interaction through the creation of high-$n$-value 2D perovskite intermediates. Given that the $n$ = 1 phase is the kinetic product of the 2D perovskite formation, it is possible that the introduction of $NH_4SCN$ slows down the crystallisation as discussed previously, hence favouring the thermodynamic sink 3D phase over the lower-$n$ phases, providing kinetic control over the resultant phase for the FPEA perovskite. Given that this slowing effect is less significant for the FBZA perovskite, we suggest that in this case it is not sufficient to effectively suppress the $n$ = 1 phase. This may be further impacted by the fact that the pristine FBZA sample does not appear to contain $n$ = 3 - 4 phases, suggesting that these phases may be thermodynamically unstable. Given that intermediate phases can act as intermediates between low-$n$ and 3D phases, this may prevent the gradual transformation of phases[53].

Absolute photoluminescence (PL) spectra of the perovskites are shown in **Figure 2c-d**. For all perovskites, irrespective of cation or additive, emission arises exclusively from the 3D phase between 700 and 800 nm, with no peaks occurring at the shorter wavelengths corresponding to excitonic behaviour as observed in the absorption despite the laser wavelength being sufficient to excite these lower-$n$ phases. As the absorption spectra shows clear evidence of the presence of these lower $n$ phases in the film, we infer that photogenerated carriers within these phases must funnel into the narrower bandgap 3D domains before recombining. For both the FPEA and FBZA perovskites, the PL is redshifted upon the addition of the additive. This shift is approximately 15 nm for FPEA and 20 nm for FBZA, aligning with the in-situ PL shown in **Figure 1**. Though both systems exhibit a red-shifting of PL, in the FPEA system this is a 'true' shift of the emission centre, while for FBZA the shift arises due to a suppression of the high-energy tail, with the low-energy PL onset remaining anchored. This suggests that for FBZA this shift arises primarily due to a passivation effect enabling more efficient transfer to low-energy domains, while for FPEA the shift is more likely to be caused by bandgap narrowing, consistent with structural lattice changes to the 3D phase. Additionally, we observe a narrowing of the full width at half maximum (FWHM) of the PL peak with the addition of the additive, from (50.9 ± 0.4) nm to (48.0 ± 0.6) nm for FPEA and (60.2 ± 0.4) nm to (51.7 ± 0.2) nm for FBZA. A narrower emission is typically indicative of a reduction in disorder within the crystal lattice, suggesting that the 3D phase has improved crystallinity with the addition of $NH_4SCN$.

We further measured the photoluminescence quantum yield (PLQY) of the samples. For each condition two samples were measured at three different locations on the substrate, giving a sample size of six. The results are shown in **Figure 2e**. We performed independent t-tests ($\alpha$ = 0.05) to determine the statistical significance of the change in PLQY with the addition of the additive. For FPEA, the PLQY significantly increased with the addition of $NH_4SCN$ ($t(5) = 7.2$, $p < .001$), with the mean PLQY increasing from (0.6 ± 0.1)% to (1.1 ± 0.2)%. The FBZA perovskite, however, demonstrated no significant change in PLQY with the addition of the additive ($t(5) = -1.5$, $p = .1$) despite higher mean PLQYs of (1.4 ± 0.3)% and (1.2 ± 0.2)% for pristine and $NH_4SCN$-treated samples respectively. The higher PLQY of FBZA vs FPEA samples (both pristine and $NH_4SCN$-treated) can be explained by the absorption spectra in **Figure 2a-b**. The presence of multiple phases ($n$ = 1-4 or 2-4) in the FPEA systems suggests that there are more interfaces between structures in these samples, compared with FBZA where there are comparably fewer distinct phases. The presence of multiple interfaces increases trap state density and non-radiative recombination, reducing PLQY.

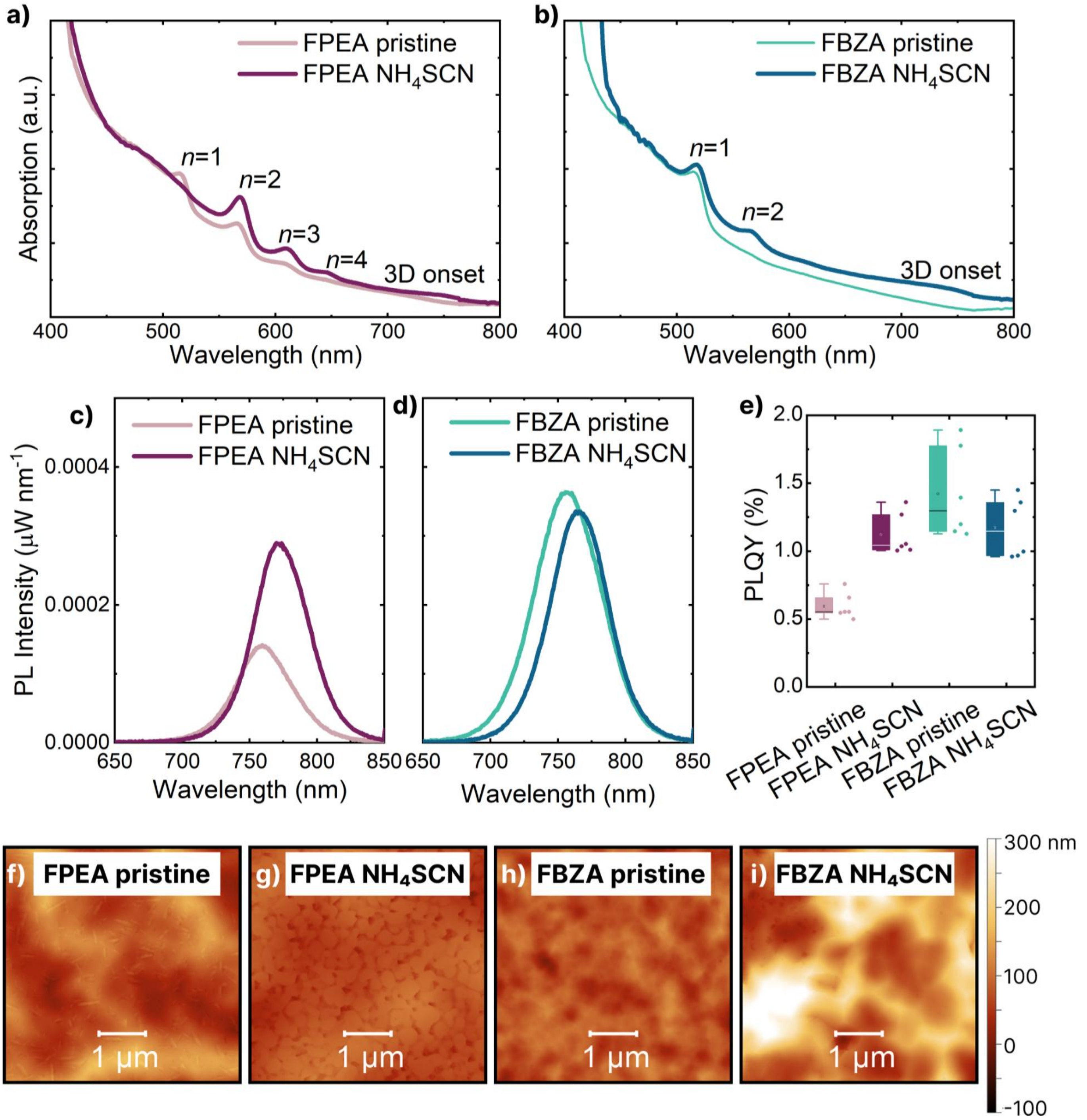

**Figure 2:** a-b) Absorption coefficients for a) $FPEA_2MA_3Pb_4I_{13}$ films with (dark pink) and without (pink) $NH_4SCN$ and b) $FBZA_2MA_3Pb_4I_{13}$ films with (dark blue) and without (light blue) $NH_4SCN$. Features corresponding to 2D and 3D phases are labelled. c-d) Representative photoluminescence spectra for each sample. e) Photoluminescence quantum yields (PLQY) for each sample. Two substrates per sample were measured with three spots per substrate. Square represents the mean, horizontal line represents the median, and whiskers represent the range between one interquartile range. f-i) 5x5 μm atomic force microscopy (AFM) images for f) pristine FPEA, g) $NH_4SCN$ FPEA, h) pristine FBZA and i) $NH_4SCN$ FBZA films.

We performed PLQY measurements for pristine and $NH_4SCN$-treated $MAPbI_3$ samples, in order to uncover the effect of the additive on the 3D phase without the influence of the 2D spacer cation. The spectra (shown in **Figure S3**) show no redshift upon the addition of $NH_4SCN$, meaning that the PL shift observed in the quasi-2D systems cannot arise due to $SCN^-$ passivation of the $MAPbI_3$ perovskite. This demonstrates that this shift must be induced by an interaction of the spacer cation with the additive, likely influenced by the presence of quasi-2D phases. Furthermore, the PLQY values did not significantly increase with the addition of $NH_4SCN$, with the pristine sample having an average PLQY of (4 ± 1)% and the $NH_4SCN$-treated sample having a mean PLQY of (3 ± 1)% ($t(5) = 1.6$, $p = 0.1$). This therefore suggests that the increase in PLQY in the FPEA sample cannot arise exclusively from a simple passivation of the 3D phase by the additive. Similarly to FBZA, the higher average PLQY of $MAPbI_3$ is likely due to the presence of only one distinct crystallographic phase in the film, therefore reducing the ubiquity of interfaces in the film.

We performed time-resolved photoluminescence spectroscopy (TRPL), with a 460 nm laser, exciting carriers in both the 3D and 2D regions in the film, and collected the integrated emission centred at 750 nm, i.e., the emission from the 3D phase. The PL kinetics are shown in **Figure S4**. In both the FPEA and FBZA perovskites, the decay is elongated with the addition of $NH_4SCN$, indicating longer charge-carrier lifetimes with the additive. This can arise due to a reduction in defects in the 3D perovskite domains and may also indicate a reduction in effective radiative recombination rates resulting from the suppressed formation of low-dimensional domains. This effect can result from the reduction of low-*n* phases where rapid excitonic recombination occurs, and also from the increased size of 3D-like domains where the main PL emission originates from. In the FPEA case, the emission shows a distinct rise soon after the initial laser pulse, suggesting that carriers are moving into the 3D domains of the film before decaying. This suggests a more efficient transfer of charges from 2D to 3D domains for FPEA perovskites. This could be due to the formation of high *n*-value intermediates that promote cascaded energy transfer to narrow bandgap regions.

To examine the morphological influence of the additive, we used atomic force microscopy (AFM) to image the surface of each thin film, shown in **Figure 2f-i**. The pristine FPEA film is composed of irregularly shaped grains, including needle-shaped features that are typical in low-*n* 2D perovskites[54]. The $NH_4SCN$-treated FPEA film, however, exhibits larger, more circular grains with well-defined boundaries typical of 3D perovskites. This could be indicative of a more 3D-like structure, given that the 3D perovskite lattice does not typically have a preferential growth direction[55]. In the case of FBZA, the film features are more difficult to discern from the AFM image. To overcome this, we performed integrated power spectral density (PSD) analysis on these AFM images, shown in **Figure S5**. This decomposes the surface profile into related spatial wavelengths allowing for comparison over different spatial frequency domains. Unlike a simple Fourier transform, PSD plots are normalised by bin width allowing for direct comparison of amplitudes between images. The correlation length, $\xi$, can be extracted from these plots using a k-correlation model, providing a measure of average feature size contributing to height variations[56–58]. Details of the fitting procedure are given in **Supplementary Note 1**. For the FBZA films, the correlation length (i.e. average feature size) increases with the addition of $NH_4SCN$ from 300 nm to 700 nm. Comparing these length scales to the AFM images in **Figure 2h** and **i**, this likely aligns with grain-scale features indicative of perovskite crystallites. This therefore suggests that the presence of $NH_4SCN$ increases the grain size in this system. For the FPEA perovskite films, the correlation lengths were found to be 430 nm and 185 nm for pristine and $NH_4SCN$-treated films respectively. Comparing this to **Figure 2f**, it is evident that in the pristine case, this larger correlation length aligns with the surface texture (wrinkling) present on the film. This wrinkling texture is likely a drying effect caused by rapid crystallisation[59]. This suggests that for this system, the surface roughness is dominated by large scale features rather than individual grains. Conversely, for the $NH_4SCN$-treated FPEA sample, the correlation length of 185 nm aligns with the grain-like features visible in **Figure 2g**. This suggests that this film is dominated by grain-level variation, indicating a smoother overall film. These morphological changes induced by $NH_4SCN$ addition are likely due to the slowing of crystallisation discussed previously in **Figure 1**, as fewer sites of nucleation allow crystallites to grow to larger sizes during film formation. We also calculated the root mean squared (rms) roughness from the AFM images. For FPEA perovskites, the rms roughness reduced with the addition of the additive, from 39 nm to 18 nm. For FBZA, the addition of the additive increased the rms roughness from 20 nm to 72 nm for pristine and $NH_4SCN$-treated films respectively. This could occur due to the lack of intermediate ($n > 2$) phases in these films (evidenced in **Figure 2b**), which promotes the phase-segregated growth of the $n = 1$ phase and $MAPbI_3$. As these phases grow with different morphologies, this would then contribute to surface roughness.

To probe the structural features in each film, we performed grazing-incidence wide angle X-ray scattering (GIWAXS) measurements. **Figures 3a-b** depict the azimuthally integrated profiles for measurements probing both the surface (0.3° incidence) and bulk film (2° incidence). The pristine FPEA film exhibits distinct reflections at Q = 0.38 and 1.01 $Å^{-1}$ ($d$ = 16.5 Å and 6.22 Å), which we attribute to the (*002*) reflection of the $n = 1$ 2D perovskite ($FPEA_2PbI_4$) (CCDC 1488195 ref.[60]) and the (*110*) reflection of the 3D tetragonal $MAPbI_3$ phase (CCDC 1474199 ref.[61]) respectively. The addition of ammonium thiocyanate to the FPEA perovskite significantly alters the crystal structure. The peak at 0.38 $Å^{-1}$ is extinguished, confirming the suppression of the $n = 1$ phase in the final film observed in the UV-Vis spectroscopy measurements. Conversely, despite the absorption spectra in **Figure 2** indicating the presence of intermediate ($n$ = 2 - 4) phases, these are not observed in the GIWAXS patterns.

In contrast to the FPEA case, while the pristine FBZA film exhibits a similar 3D reflection at 1.01 $Å^{-1}$ (**Figure 3b**) and (*002*) $n = 1$ (ref.[62]) reflection at 0.46 $Å^{-1}$ ($d$ = 13.7 Å), the addition of $NH_4SCN$ does not inhibit low-$n$ formation. Instead, it promotes the formation of multiple low-$n$ phases, evidenced by the persistence of the $n = 1$ reflection alongside additional peaks at 0.32, 0.63, and 0.94 $Å^{-1}$ ($d$ = 19.6 Å, 10.0 Å, and 6.68 Å). We attribute these to the (*002*) and (*004*) $n = 2$ phase and (*004*) $n = 1$ phase reflections respectively. This indicates that in the FBZA perovskite, instead of inhibiting low-$n$ phases, the addition of $NH_4SCN$ enhances the formation of $n = 1$ and $n = 2$ phases, corroborating the absorption shown in **Figure 2**.

**Figures 3c-d** show the peak position and full-width at half-maximum (FWHM) respectively for the $MAPbI_3$ (*110*) peak in all samples in the bulk film. In the FPEA films, the addition of $NH_4SCN$ shifts the position of this peak to a higher Q value by 0.016 $Å^{-1}$. The FBZA system sees a much smaller increase in peak position upon the addition of $NH_4SCN$, of only 0.007 $Å^{-1}$. The shifted peak in **Figure 3c** additionally exhibits an asymmetric shape with the addition of the additive. We suggest that this shift is therefore due to the coexistence of the original $MAPbI_3$ (*110*) peak alongside an additional peak at Q = 1.0 $Å^{-1}$. This, combined with the absorption data in **Figure 2a-b** suggesting that FPEA + $NH_4SCN$ promotes higher $n$-value phases, implies the existence of high $n$-value 2D perovskite in the FPEA + $NH_4SCN$ film which would exhibit (*111*) reflections close to Q = 1.0 $Å^{-1}$.

Further, the FWHM of the $MAPbI_3$ (*110*) peak is reduced in both quasi-2D perovskite cases, from 0.06 $Å^{-1}$ to 0.04 $Å^{-1}$ for FPEA and 0.08 $Å^{-1}$ to 0.04 $Å^{-1}$ for FBZA. A narrower peak is typically indicative of an increase in domain size, aligning with our previous results suggesting that the additive slows the crystallisation, resulting in enlarged perovskite grains. The GIWAXS results suggest that for FPEA, this enlarged grain size occurs primarily for the $MAPbI_3$ phase. However, for FBZA both the $MAPbI_3$ peak and the low-Q 2D peaks are narrowed, suggesting that the grains are enlarged for all of the phases present in the sample. The shifting of the (*110*) $MAPbI_3$ peak in the FPEA perovskite is observed primarily in the bulk, rather than on the surface of the film (**Figure S6**), indicating that high $n$-value quasi-2D perovskite is concentrated at the substrate-liquid interface whereas the surface of the film is primarily 3D perovskite.

To probe potential preferential orientation of the structures within the film, we performed azimuthal integrations across various azimuthal angles relative to the beam incidence at the detector: out-of-plane (along the $Q_z$ direction in reciprocal space) and in-plane (along the $Q_{xy}$ direction in reciprocal space), (**Figure S7**). For the FPEA perovskites (**Figure S7 a-b**), the (*110*) peak is observed in both the in- and out-of-plane integrations, with the shifting only apparent in the out-of-plane direction. This is in agreement with the suggestion that this is due to the coexistence of $MAPbI_3$ and high-$n$ quasi-2D perovskite, as the distribution of $n$-values oriented vertically with (*111*) parallel to the $Q_z$-axis would result in a shift only in the $z$-axis. As the shift does not occur isotropically, we can infer that this is not due to a simple passivating effect whereby the unit cell is modified by the intercalation of $SCN^-$ ions. Such a process would likely result in an isotropic expansion (rather than contraction, given the relative ionic radii of $I^-$ and $SCN^-$) of the unit cell. Additionally, in the FBZA perovskites the low-dimensional peaks are predominantly present in the out-of-plane direction, confirming that these phases are oriented with the 2D layers oriented parallel to the substrate, and that the addition of $NH_4SCN$ has not induced vertical orientation of these layers.

The 2D GIWAXS patterns (**Figure 3e-h**) provide additional information about the orientation of these domains in the film. The FPEA pristine film (**Figure 3e**) displays Debye-Scherrer diffraction rings, indicating polycrystalline film formation, with grains oriented randomly with respect to the substrate[63]. The addition of $NH_4SCN$ (**Figure 3f**) significantly increases the preferential orientation with respect to the substrate, evidenced by the distinct transition of features from rings to Bragg spots. In particular, the $MAPbI_3$ (*110*) reflection changes from a diffraction ring to a Bragg spot, indicating increasing preferential orientation in the $MAPbI_3$ domains. The presence of (*211*) and (*112*) features of the $MAPbI_3$ phase (marked on **Figure 3f**) at angles of 27° and 45° from the $Q_z$ axis demonstrate that the (*110*) plane is parallel to the substrate. Conversely, in the FBZA samples, neither the $MAPbI_3$ (*110*) peak nor the low-*n* peaks exhibit signs of significant changes in orientation upon addition of $NH_4SCN$, as the features present exclusively as Debye-Scherrer diffraction rings, indicating polycrystalline structure (**Figure 3g-h**). This suggests that these structural changes are unique to the FPEA system.

We additionally performed one-dimensional XRD and GIWAXS on samples of pure $MAPbI_3$, i.e., in the absence of the spacer cation. These are shown in **Figure S8**. No peak shift of the (*110*) diffraction ring was observed upon the addition of $NH_4SCN$ (**Figure S8a**) and no preferred orientation is observed (**Figure S8b-c)**, indicating that both of these effects are dependent on the presence of the spacer cation. The diffraction rings are narrowed, suggesting larger crystalline domains, however features remain as Debye-Scherrer diffraction rings rather than Bragg spots. This suggests that while the role of $NH_4SCN$ in increasing grain size may be independent to the quasi-2D phases, the additive alone (in the absence of quasi-2D phases) is not responsible for the compression of the $MAPbI_3$ unit cell nor the oriented 3D perovskite growth observed in the FPEA quasi-2D perovskite system. Along with **Figure S3** demonstrating that the addition of pure $NH_4SCN$ to pure $MAPbI_3$ does not result in red shifting of the PL emission, these results unambiguously confirm that the role of $NH_4SCN$ in quasi-2D perovskites is beyond that of a simple passivation. The structural data suggest that the origin of the redshifted PL emission arises from the introduction of high-*n* quasi-2D domains, which promote funnelling into the narrow bandgap 3D regions, for example by Förster resonance energy transfer[64].

To provide a structural basis for these findings, we simulated[65] the GIWAXS pattern of both $MAPbI_3$ and $n$ = 2 FPEA perovskite aligning vertically and horizontally with respect to the substrate. The simulated patterns are shown **in Figure S9** alongside graphical illustrations of the respective crystal orientations. The (*110*), (*211*), (*112*) and (*1-10*) peaks in **Figure 3f** align almost exactly with the simulated pattern in **Figure S9a** corresponding to the pure $MAPbI_3$ phase oriented with the (*110*) plane parallel to the substrate. Additionally, the simulated $n$ = 2 FPEA perovskite GIWAXS patterns display no peaks that can be identified in our data, in neither a vertical **(Figure S9b**) nor horizontal (**Figure S9c**) orientation. In particular, low-Q (*002*) peaks are absent in both $Q_z$ and $Q_{xy}$. If low-*n* domains were present in the film, these features would be expected to have the most intense scattering, as shown in **Figure S9b-c**. Hence, their absence implies that any low-*n* domains are not crystalline enough to scatter perceptibly. This confirms that the increased ordering seen in our samples is the result of highly (*110*)-oriented $MAPbI_3$ perovskite, not oriented low-*n* 2D layers and hence we show that the 3D phase is the dominant crystal structure in these samples.

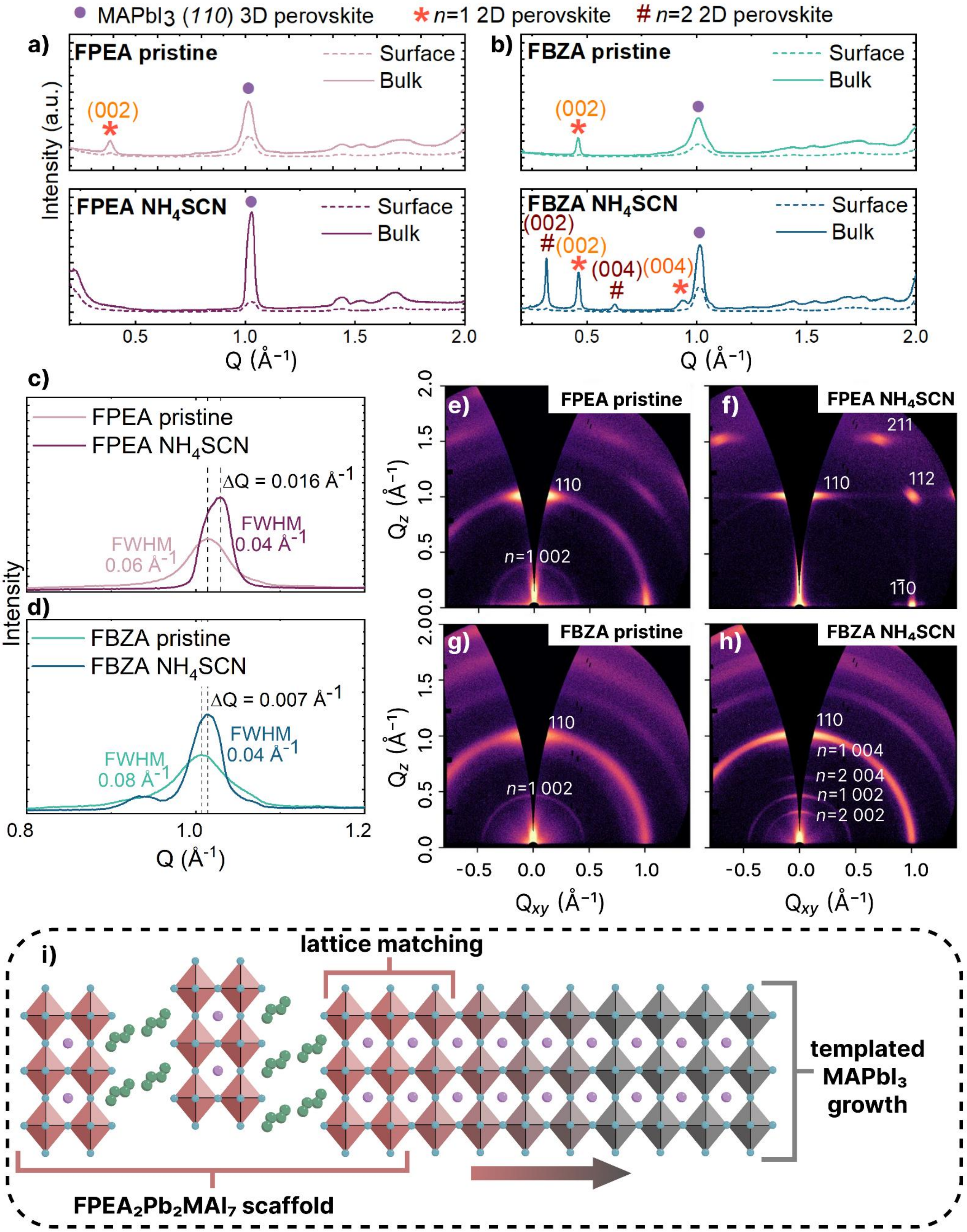

**Figure 3:** a-b) Azimuthally integrated grazing incidence wide angle X-ray scattering (GIWAXS) profiles of a) $FPEA_2MA_3Pb_4I_{13}$ films with (dark pink) and without (light pink) $NH_4SCN$ and b) $FBZA_2MA_3Pb_4I_{13}$ films with (dark blue) and without (light blue) $NH_4SCN$ for both the surface (0.3° incidence angle, dashed line) and bulk (2° incidence angle, solid line) of the film. Characteristic peaks of 2D and 3D phases are marked with symbols. c-d) Zoom-in of the $MAPbI_3$ (*110*) peak in each film. Dashed lines represent the peak position. e-h) GIWAXS reciprocal space colourmaps of FPEA perovskites without (e) and with (f) $NH_4SCN$ and FBZA perovskites without (g) and with (h) $NH_4SCN$. Labelled features correspond to $MAPbI_3$ structure unless otherwise labelled. i) Schematic of the proposed templating procedure for $MAPbI_3$ film growth.

We additionally observe a horizontal 'streaking' at Q = 1 $Å^{-1}$ in the additive-treated FPEA GIWAXS profile in **Figure 3f**. This aligns with the presence of high-*n* quasi-2D phases in this sample, providing a graded phase distribution between 2D and 3D phases. The (*111*) peaks of high-*n* phases lie at $Q_z \sim 1$ $Å^{-1}$, extending horizontally from the $Q_z$ axis when these regions are oriented vertically with (*111*) parallel to $Q_z$. With increasing n these peaks lie correspondingly closer to the axis, therefore in the presence of a distribution of high-*n* phases, a horizontal line is observed. This confirms that in the FPEA + $NH_4SCN$ film there exists both highly ordered $MAPbI_3$ and a distribution of high-*n* 2D perovskite phases. The lack of (*002*) reflections discussed above can be explained by these reflections sitting at very low Q. Though detailed crystallographic information about such phases is limited due to the inability to synthesise high-*n* single crystals, we can approximate the lattice parameters by adding subsequent octahedral spacing distances (of ~ 8.8 Å) along the *c*-axis. This gives an approximate *c*- parameter of ~115 Å for *n* = 10, or Q = 0.1 $Å^{-1}$, which would further decrease for increasing *n*. While these peaks may be present, the limitations of the detector mean that for Q values near or below 0.1 $Å^{-1}$, the beamstop and related interference obscure features in this region. Additionally, poor phase purity (associated with a distribution of high-*n* phases, rather than one distinct *n*-value) results in poor crystallinity in the *c*-axis, therefore reducing the intensity of (*002*) reflections in any case.

The addition of $NH_4SCN$ does not result in these structural changes or enhanced ordering for the FBZA quasi-2D perovskite, and so it is clear that the impact of this additive is highly dependent on the spacer cation. To probe this further, we replaced FPEA with unsubstituted phenethylammonium (PEA) and measured the GIWAXS profiles both with and without $NH_4SCN$. These are depicted in **Figure S10**. Similar to the FPEA system, the PEA-based quasi-2D perovskite exhibits highly oriented $MAPbI_3$ perovskite, with the (*110*) plane parallel to the substrate. Also mirroring the FPEA system, the PEA samples show distinct suppression of the *n* = 1 phase, evidenced by the extinction of the (*002*) peak at ~ 0.45 $Å^{-1}$. Hence, we conclude that $NH_4SCN$ promotes the growth of highly oriented crystalline $MAPbI_3$ in the presence FPEA and PEA, but not FBZA spacer cations. We suggest that this is partly due to the energetic unfavourability of high-*n* quasi-2D phases in the FBZA system. This is evidenced in the UV-Visible absorption (**Figure 2b**) where phases *n* > 2 are not present. This also explains the strong diffraction of low-*n* peaks in these samples, as there is more organic available to form large domains of low-*n* material.

In the case of FPEA and PEA we suggest then that the mechanism behind the changing structure with $NH_4SCN$ is structural templating due to lattice matching between the (*110*) $MAPbI_3$ plane and a specific quasi-2D phase, illustrated in **Figure 3i**. The addition of $NH_4SCN$ promotes the formation of high-*n* quasi-2D phases which subsequently induce in-*situ* structural templating, forming highly ordered $MAPbI_3$. Using single crystal structures from the literature, we aligned the lattice parameters of the (*110*) peak (6.12 Å) with that of the *n* = 2 phase of FPEA ($FPEA_2MAPb_2I_7$, CCDC 1483085 ref.[60]) (*110*) lattice plane (6.131 Å) and PEA ($PEA_2MAPb_2I_7$, ref[66]). Though single crystal structures are not available for equivalent structures with *n* > 2, we infer that the (*110*) lattice plane would remain largely similar for *n* >= 2, since additional octahedral layers primarily affect the *c*-axis. This confirms that these structures possess the correct lattice parameters to template the $MAPbI_3$ perovskite, illustrated with the dashed line in **Figure S7**. Given that $NH_4SCN$ influences both the templating effect and the phase formation in the FPEA and PEA perovskites, we suggest that these two effects happen concomitantly with the *n* = 1 suppression being necessary for the templating. Given the low enthalpy of formation for the *n* = 1 phase, we suggest that when this phase is present the rapid crystallisation interrupts the growth of the $MAPbI_3$ phase and therefore inhibits templating. The requirements for the spacer cation to enable templated growth are therefore twofold: the additive must effectively inhibit the *n* = 1 phase and promote high-*n* quasi-2D phases, and the lattice parameters must align to enable lattice matching of the target 3D phase. This is further evidenced by the fact that we do not observe templating in the FBZA perovskite. Though the FBZA *n* = 2 (*110*) lattice parameter (6.19 Å ref.[67]) suggests that this perovskite may be able to template the $MAPbI_3$ phase, the persistence of the *n* = 1 phase, and the lack of high-*n* phases in this perovskite prevents templating from occurring.

As suppression of the *n* = 1 phase is critical to the templating effect, understanding this mechanism is necessary. To study the film formation kinetics, we measured in-*situ* XRD during annealing of $NH_4SCN$-treated samples to determine the degree of DMF coordination during annealing**. Figure S11** shows the $PbI_2$ · DMF solvate intermediate XRD peaks for each sample, situated around 9.9° and 10.2°. The FBZA sample is unique in that these features are more intense than in the FPEA, PEA, or $MAPbI_3$ samples and they do not diminish during annealing. This suggests that DMF is more strongly coordinated in the FBZA system, which is likely responsible for the slow PL evolution shown in **Figure 1c**. Compared to FPEA, FBZA differs only in a reduced alkyl chain length, which will change the local electrostatic potential and likely acts to stabilize the solvate forms. Given that $SCN^-$ is known to coordinate to $Pb^{2+}$ (refs.[35,68]) we suggest that the stable solvate phase in the FBZA sample inhibits the coordination of $SCN^-$ to $Pb^{2+}$, which prevents the inhibition of the *n* = 1 phase and the templating effect.

To understand the phase evolution in these samples, we placed $NH_4SCN$-treated samples in a humidity chamber and studied the quasi-2D phase evolution. The presence of humidity acts to facilitate mass transfer within the film and evolve the system towards its equilibrium phase distribution[69]. The XRD profiles over time are shown **in Figure S12**. For both FPEA and PEA quasi-2D perovskites, features corresponding to *n* = 2, 3, and 4 are clearly visible, with *n* = 2 and 3 being the most prominent. Though the intermediate 2D phases grow in intensity over time, the *n* = 1 phase is not present even after aging. As it has previously shown that intermediate phases would transform to low-*n* or *n* = 1 phases in the presence of moisture[69,70], this further demonstrates the efficient suppression of the *n* = 1 phase and relative stability of the intermediate phases. This data also structurally confirms the presence of the *n* = 2 phase necessary for templating. For the FBZA perovskite, though quasi-2D *n* = 2 and 3 phases are initially visible in the XRD profile, they degrade rapidly in ~ 5 minutes, with a large FBZA · $H_2O$ complex becoming prominent and only the *n* = 1 phase persisting. This suggests that FBZA perovskite does not form stable high-*n* quasi-2D perovskite phases, and/or that these phases are not thermodynamically favourable, further confirming that templating is not possible in this system. We suggest that the unfavourability of high-*n* phases in this system could be due to steric hindrance, with the shorter alkyl chain of FBZA preventing the ammonium head from sitting sufficiently deep into the octahedral pocket. This would introduce severe octahedral tilting and lattice strain, resulting in chronic instability.

The disparity between the prominent *n* = 2, 3, and 4 excitonic peaks in UV-Vis absorption (**Figure 2a-b**) and the absence of corresponding low-Q diffraction reflections in GIWAXS demonstrates that the composition of the $NH_4SCN$-treated films is largely that of 3D perovskite. While optical absorption detects localized quantum confinement within individual or few-layer organic–inorganic sheets, GIWAXS requires long-range, out-of-plane translational stacking order across multiple periodic unit cells to generate visible Bragg peaks. In freshly prepared FPEA samples with NH4SCN, these quasi-2D intermediate phases exist as ultra-small, isolated, or translationally disordered templates embedded within the bulk. Upon controlled ambient exposure in a humidity chamber, slow moisture-assisted recrystallization facilitates the gradual coalescence of these domains, driving the progressive emergence and sharpening of low-Q scattering features during film aging. This highlights the fact that the role of $NH_4SCN$ lies in improving quasi-2D perovskite photovoltaic properties lies primarily in the growth of highly crystalline, templated 3D perovskite, rather than oriented low-*n* quasi-2D perovskite layers as has been previously suggested.

The use of 2D perovskite structures to template the growth of 3D perovskite has been demonstrated previously[71–75]. These studies target the scaffolding 2D phase by pre-synthesizing a single crystal and adding this into the precursor solution, or by using an initial 2D thin film to encourage templated growth. In this work, we have shown that by sufficiently controlling the phase formation through the use of additives, templated growth can be induced in-*situ* without necessitating single crystal growth of the target phase. Moreover, we have shown that a graded distribution of *n*-values can instigate templating provided there is sufficient lattice matching between phases. Notably, templating can occur for a graded distribution of *n*-values as the (*110*) lattice plane is mostly unchanged with varying *n*, with the exception of *n* = 1 due to the absence of the cage cation. The *n* = 1 phase therefore must be supressed to enable templating, likely because lower-*n* phases are kinetically favoured over higher-*n* phases[76], meaning lower-*n* phases have larger crystal domains and rapid crystallisation that can more easily disrupt growth. This is evident in the fact that only the $n = 1$ phase is visible in XRD patterns despite other phases clearly being visible in absorption spectra. The role of $NH_4SCN$ is therefore to extend the nucleation window, allowing the $n = 1$ kinetic trap to be bypassed. $SCN^-$ incorporation has been shown to have a similar effect in delaying nucleation in multiple perovskite systems[77–79].

To measure the impact of this templated growth on perovskite photovoltaics, we fabricated *p-i-n* solar cells with a device architecture of indium tin oxide (ITO) / (2-(3,6-Dimethoxy-9H-carbazol-9-yl)ethyl)phosphonic acid (MeO-2PACz) / perovskite / $C_{60}$ / bathocuproine (BCP) / Ag. **Figure 4a-b** shows the current density - voltage (*JV*) curves for FPEA and FBZA 2D perovskites respectively, with and without $NH_4SCN$. The effect of the additive on the shape of the *JV* curve is most apparent in the FPEA case, where both the $J_{SC}$ and shunt resistance, $R_{SH}$, are increased. For FBZA, the $J_{SC}$ is also increased but to a lesser extent. The performance metrics are displayed in **Figure 4c-f** for 32 devices per condition (forward and reverse scan data points are plotted separately). The $NH_4SCN$-treated FPEA devices exhibit a short-circuit current ($J_{SC}$) of (17.1 ± 1.0) mA $cm^{-2}$, an open-circuit voltage ($V_{OC}$) of (1.09 ± 0.02) V and a fill factor (FF) of (70 ± 6)%, yielding a mean power conversion efficiency (PCE) of (13.2 ± 1.5)%. In comparison, the pristine FPEA devices yield a mean PCE of (4.6 ± 1.0)%, largely due to the much lower $J_{SC}$ and FF of (9.2 ± 2.0) mA $cm^{-2}$ and (45 ± 3)%.

The device performance parameters are summarised in **Table 1**. We performed a Welch-corrected t-test ($\alpha = 0.05$) to compare the mean PCE between the pristine and $NH_4SCN$-treated FPEA samples, which showed that the $NH_4SCN$-treated samples yielded a significantly higher PCE ($t(82) = -31$, $p < .001$). The increase in $J_{SC}$ can be ascribed to the improved crystallinity and increased grain size in the templated $MAPbI_3$, which reduces the number of trap states and grain boundaries that prevent charges from being extracted from the perovskite. Additionally, the inhibition of the $n = 1$ phase also acts to increase the $J_{SC}$, as this wide bandgap phase can act as a charge blocking layer and prevent the extraction of charges. The additive also increases the external quantum efficiency (EQE) (**Figure 4g**) across all wavelengths in both perovskites, and particularly in the FPEA case the EQE is increased in the 400 - 550 nm range. This could be due to the inhibition of the $n = 1$ phase which absorbs at 520 nm. The integrated current as derived from the EQE spectra matches the $J_{SC}$ of devices to within 5%. The increase in fill factor for the $NH_4SCN$-treated sample is likely due to the increased shunt resistance, $R_{SH}$, as is clear in the *JV* curve in **Figure 4a**. This is suggestive of a reduction in crystal defects due to the templated perovskite being highly crystalline[43]. We additionally calculated Urbach energies ($E_u$) for each perovskite using the EQE spectra. These are shown in **Figure S13**. Regardless of additive, the FBZA perovskites have larger $E_u$, reaching the order of the thermal energy at room temperature (21 meV and 20 meV for pristine and $NH_4SCN$-treated FBZA samples respectively, compared with $k_BT$ ~ 26 meV at 300 K). This points to an increase in sub-gap absorption in these perovskites, potentially due to low crystallinity and high defect density. Conversely, FPEA perovskites have lower $E_u$ values of 18 meV and 17 meV for pristine and $NH_4SCN$-treated perovskites respectively[80].

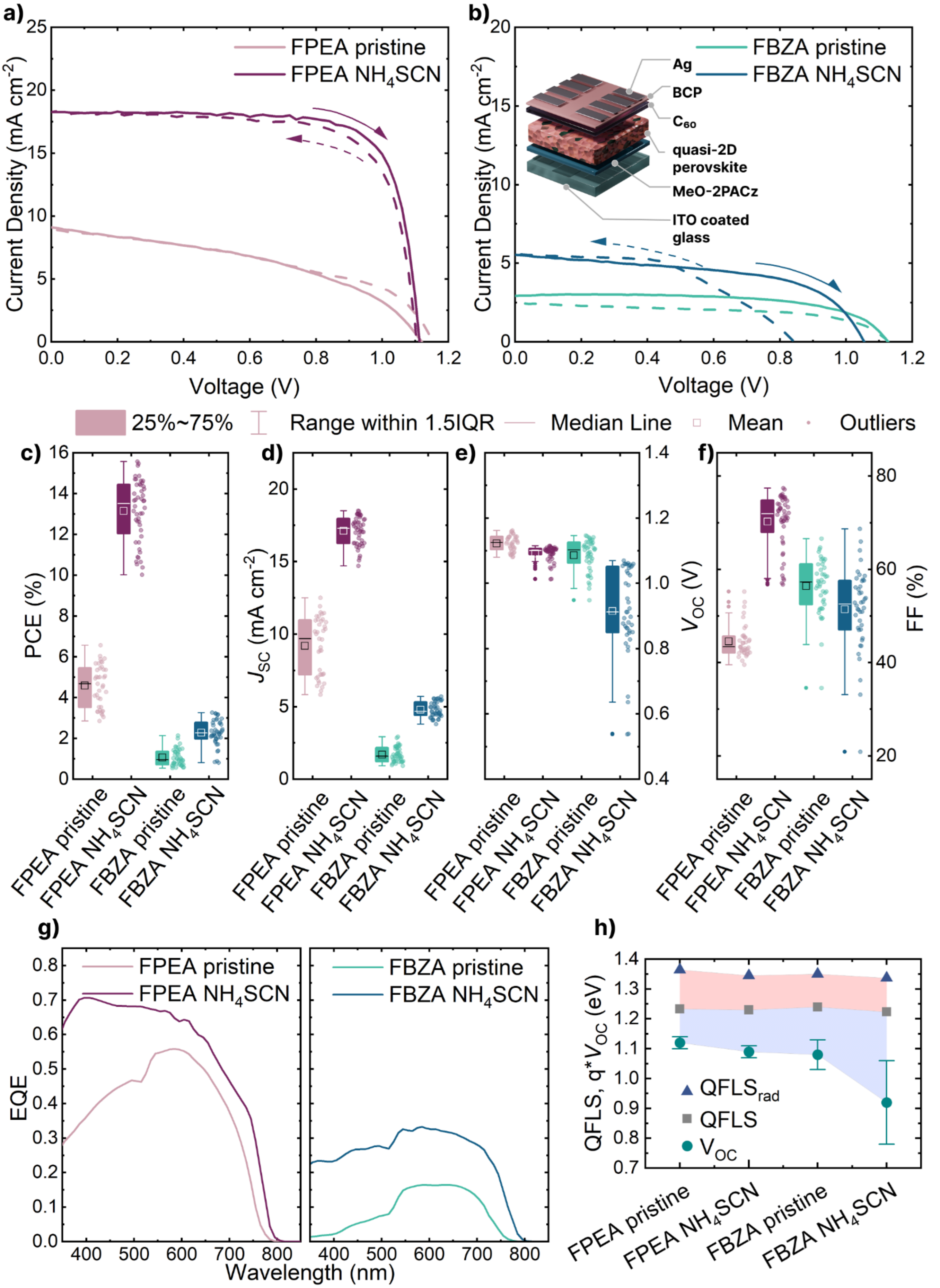

**Figure 4:** a-b) Current density - Voltage (*JV*) characteristics for a) $FPEA_2MA_3Pb_4I_{13}$ and b) $FBZA_2MA_3Pb_4I_{13}$ perovskite devices with and without $NH_4SCN$ with an active area of 0.02 $cm^2$. Inset shows schematic of device architecture. Arrows indicate forward and reverse voltage sweep directions. c-f) Box plots showing device performance metrics for each condition: c) power conversion efficiency (PCE), d) short circuit current ($J_{SC}$) e) open circuit voltage ($V_{OC}$) and f) fill factor (FF). g) Photovoltaic external quantum efficiency (EQE) for the champion device from each condition. h) Mean $V_{OC}$ (green dot, whiskers represent one standard deviation), radiative quasi-Fermi level splitting ($QFLS_{rad}$, blue triangle) and QFLS (grey square) for each condition.

**Table 1:** Champion (mean ± one standard deviation) photovoltaic performance metrics for each perovskite condition.

| | PCE (%) | $J_{SC}$ (mA cm$^{-2}$) | $V_{OC}$ (V) | FF (%) |
|---|---|---|---|---|
| **Pristine FPEA** | 6.6 (4.6 ± 1.0) | 12.5 (9.2 ± 2.0) | 1.16 (1.12 ± 0.02) | 55 (45 ± 3) |
| **FPEA $NH_4SCN$** | 15.6 (13.2 ± 1.5) | 18.5 (17.1 ± 1.0) | 1.12 (1.09 ± 0.02) | 77 (70 ± 6) |
| **Pristine FBZA** | 2.1 (1.1 ± 0.4) | 2.9 (1.7 ± 0.5) | 1.15 (1.08 ± 0.05) | 67 (56 ± 6) |
| **FBZA $NH_4SCN$** | 3.3 (2.3 ± 0.6) | 5.7 (4.8 ± 0.5) | 1.1 (0.92 ± 0.14) | 69 (51 ± 9) |

For FBZA perovskites, the photovoltaic performance is significantly lower than FPEA with the PCE being (1.1 ± 0.4)% and (2.3 ± 0.6)% for pristine and $NH_4SCN$-treated samples respectively. Although this is still a significant increase ($t(69) = -10.6$, $p < .001$), the performance is drastically inhibited by the low $J_{SC}$ of (1.7 ± 0.5) mA cm$^{-2}$ for pristine and (4.8 ± 0.5) mA cm$^{-2}$ for $NH_4SCN$-treated FBZA perovskites. This could arise due to the high prominence of the $n = 1$ phase in this sample which acts as a charge blocking layer and a site for non-radiative recombination.

In both perovskites, the $V_{OC}$ is reduced with the addition of $NH_4SCN$. We attribute this to a narrowing of the bandgap caused by the additive, evidenced both in the red shifting of PL shown in **Figure 2** and the redshift of the band edge in the external quantum efficiency (EQE) spectra (from 745 nm to 775 nm for FPEA perovskites and from 725 nm to 755 nm for FBZA perovskites) shown in **Figure 4g**. This corresponds to a shifting of the photovoltaic bandgap from 1.66 eV to 1.60 eV in the FPEA case with the addition of $NH_4SCN$ (**Figure S14**). For FBZA perovskites, this shift is from 1.71 eV to 1.64 eV. To understand the severity of the $V_{OC}$ reduction from the thermodynamic limits of these materials at their respective bandgaps, we used the EQE spectra (**Figure 4g**) to calculate the quasi-Fermi level splitting (QFLS). Full details are provided in **Supplementary Note 2**. The QFLS can be considered the 'best case $V_{OC}$' and therefore we depict it alongside the $V_{OC}$ in **Figure 4h,** in both the radiative limit ($QFLS_{rad}$, triangles) and when considering non-radiative losses as assessed through PLQY measurements (squares). The pink shaded area represents the non-radiative losses between $QFLS_{rad}$ and QFLS, which are largest for the pristine FPEA sample. We attribute this to heterogeneity introducing sites for non-radiative recombination, evidenced in our morphological measurements in **Figure 2f** and the broad phase distribution observed in our absorption measurements (**Figure 2a**). The blue shaded area denotes the mismatch between the QFLS and the achieved $V_{OC}$, often attributed to energetic misalignment within the device stack[81–83]. This loss mechanism is greatest for the FBZA $NH_4SCN$ sample, which we suggest is owing to the ubiquity of wide bandgap regions of low-*n* perovskite as evidenced in our spectroscopic and structural measurements.

It is evident that the effect of $NH_4SCN$ on quasi-2D perovskite devices is highly dependent on the choice of spacer cation. For FPEA perovskites, device performance is significantly improved, largely *via* an increase in $J_{SC}$ and FF. Conversely for FBZA, both $J_{SC}$ and FF remain limited in both the pristine and additive-treated cases. This can be explained by the templating mechanism discussed previously: in the FPEA case, where we observe templated growth, we attribute the increase in $J_{SC}$ to a reduction in low-*n* 2D phases (which inhibit charge transport) as well as an increase in crystallinity of the 3D perovskite phase, increasing charge carrier mobility. Similarly, the increased FF can be explained by the improved morphology (evidenced in our morphological measurements in **Figure 2**) and increased $R_{SH}$ which again is indicative of a more crystalline perovskite. In the FBZA case where we do not observe templating, we do not see these performance enhancements. This is particularly interesting considering that the molecular structures of FPEA and FBZA differ by only a single carbon in the alkyl chain.

## Conclusions:

We have demonstrated that the beneficial effect of $NH_4SCN$ on charge-carrier transport in quasi-2D perovskites is highly dependent on the choice of spacer cation. Contrary to previous reports where $NH_4SCN$ was suggested to vertically orient low-*n* phases of FPEA quasi-2D perovskites, we have shown that the inclusion of $NH_4SCN$ in the perovskite precursor solution promotes an *in-situ* templated growth mechanism of 3D $MAPbI_3$ utilising the quasi-2D phases as a structural template. This templated growth improves crystallinity, morphology, and optoelectronic properties increasing photovoltaic performance. This is evidenced by a clear increase in crystallographic ordering of the $MAPbI_3$ perovskite, apparent in our GIWAXS studies. Key to this templating mechanism is the inhibition of the $n$ = 1 2D phase and promotion of higher-*n* phases, induced by the $NH_4SCN$ addition which provides kinetic control over the 2D perovskite phase formation.

Conversely, we show that structural templating effects are not observed in systems based on similar cations (FBZA, FPA). In FBZA perovskites we demonstrate that the persistence of the kinetically favoured $n$ = 1 phase and energetic unfavourability of higher-*n* phases in this material inhibit templated growth, thereby underpinning the necessity for both kinetic control and structural compatibility in enabling templating. These results provide critical insights into the use of additives in quasi-2D perovskites highlighting the importance of the spacer cation in determining their impact.

## References:

1. Zhao X, Liu T, Loo YL. Advancing 2D Perovskites for Efficient and Stable Solar Cells: Challenges and Opportunities. *Adv Mater*. 2022;34(3):2105849. doi:10.1002/adma.202105849

2. Chen P, Bai Y, Wang S, Lyu M, Yun JH, Wang L. In Situ Growth of 2D Perovskite Capping Layer for Stable and Efficient Perovskite Solar Cells. *Adv Funct Mater*. 2018;28(17):1706923. doi:10.1002/adfm.201706923

3. Leung TL, Ahmad I, Syed AA, Ng AMC, Popović J, Djurišić AB. Stability of 2D and quasi-2D perovskite materials and devices. *Commun Mater*. 2022;3(1):1-10. doi:10.1038/s43246-022-00285-9

4. Tsai H, Nie W, Blancon JC, et al. High-efficiency two-dimensional Ruddlesden–Popper perovskite solar cells. *Nature*. 2016;536(7616):7616. doi:10.1038/nature18306

5. Li Z, Yang M, Park JS, Wei SH, Berry JJ, Zhu K. Stabilizing Perovskite Structures by Tuning Tolerance Factor: Formation of Formamidinium and Cesium Lead Iodide Solid-State Alloys. *Chem Mater*. 2016;28(1):284-292. doi:10.1021/acs.chemmater.5b04107

6. Mao L, Teicher SML, Stoumpos CC, et al. Chemical and Structural Diversity of Hybrid Layered Double Perovskite Halides. *J Am Chem Soc*. 2019;141(48):19099-19109. doi:10.1021/jacs.9b09945

7. Lee H, Moon T, Lee Y, Kim J. Structural Mechanisms of Quasi-2D Perovskites for Next-Generation Photovoltaics. *Nano-Micro Lett*. 2025;17(1):139. doi:10.1007/s40820-024-01609-9

8. Katan C, Mercier N, Even J. Quantum and Dielectric Confinement Effects in Lower-Dimensional Hybrid Perovskite Semiconductors. *Chem Rev*. 2019;119(5):3140-3192. doi:10.1021/acs.chemrev.8b00417

9. Lee S, Cho H, Kang S, Oh OJ, Kim DH, Noh JH. Deciphering 2D perovskite's role in perovskite solar cells via intact 3D/2D junctions. *Energy Environ Sci*. 2024;17(17):6234-6244. doi:10.1039/D3EE04520J

10. Yang L, Li X, Li M, Liu D, Wang D, Wang T. Progress of 2D Perovskite Solar Cells: Structure and Performance. *Small*. 2025;21(7):2410731. doi:10.1002/smll.202410731

11. Zhang L, Sun C, He T, et al. High-performance quasi-2D perovskite light-emitting diodes: from materials to devices. *Light Sci Appl*. 2021;10(1):61. doi:10.1038/s41377-021-00501-0

12. Tyagi D, Laxmi V, Basu N, et al. Recent advances in two-dimensional perovskite materials for light-emitting diodes. *Discov Nano*. 2024;19(1):109. doi:10.1186/s11671-024-04044-2

13. Wang S, Ling Z, Blom PWM, Marszalek T, Pisula W. Engineering Strategies for 2D Layered Tin Halide Perovskite Field-Effect Transistors. *Adv Funct Mater*. 2025;35(48):2501217. doi:10.1002/adfm.202501217

14. Kumavat SR, Sachdeva G, Sonvane Y, Gupta SK. Structural and compositional properties of 2D CH3NH3PbI3 hybrid halide perovskite: a DFT study. *RSC Adv*. 2022;12(40):25924-25931. doi:10.1039/D2RA02874C

15. Pandey M, Jacobsen KW, Thygesen KS. Band Gap Tuning and Defect Tolerance of Atomically Thin Two-Dimensional Organic–Inorganic Halide Perovskites. *J Phys Chem Lett*. 2016;7(21):4346-4352. doi:10.1021/acs.jpclett.6b01998

16. Meheretu GM, Worku AK, Yihunie MT, Koech RK, Wubetu GA. The recent advancement of outdoor performance of perovskite photovoltaic cells technology. *Heliyon*. 2024;10(17):e36710. doi:10.1016/j.heliyon.2024.e36710

17. Qu W, Zong J, Feng X, Song J, Yang B, Wei H. Preferred Orientation and Phase Distribution of Quasi-2D Perovskite for Bifunctional Light-Emitting Photodetectors. *Nano Lett*. 2024;24(25):7593-7600. doi:10.1021/acs.nanolett.4c00927

18. Liang C, Gu H, Xia Y, et al. Two-dimensional Ruddlesden–Popper layered perovskite solar cells based on phase-pure thin films. *Nat Energy*. 2021;6(1):1. doi:10.1038/s41560-020-00721-5

19. Stippell E, Li W, Quarti C, Beljonne D, Prezhdo OV. Enhancing Interlayer Charge Transport of Two-Dimensional Perovskites by Structural Stabilization via Fluorine Substitution. *ACS Appl Mater Interfaces*. 2024;17(1):2032-2040. doi:10.1021/acsami.4c17876

20. Zhang Y, Abdi-Jalebi M, Larson BW, Zhang F. What Matters for the Charge Transport of 2D Perovskites? *Adv Mater*. 2024;36(31):2404517. doi:10.1002/adma.202404517

21. Kim K, Kwon N, Park SW, et al. Exciton Binding Energy Modulation in 2D Perovskites: A Phenomenological Keldysh Framework. *Adv Funct Mater*. n/a(n/a):e20461. doi:10.1002/adfm.202520461

22. Niedzwiedzki DM, Zhou H, Biswas P. Exciton Binding Energy of MAPbI3 Thin Film Elucidated via Analysis and Modeling of Perovskite Absorption and Photoluminescence Properties Using Various Methodologies. *J Phys Chem C*. 2022;126(2):1046-1054. doi:10.1021/acs.jpcc.1c09598

23. Even J, Pedesseau L, Katan C. Understanding Quantum Confinement of Charge Carriers in Layered 2D Hybrid Perovskites. *ChemPhysChem*. 2014;15(17):3733-3741. doi:10.1002/cphc.201402428

24. Lee YH, Macam GM, Villaos RAB, et al. Effect of MA Orientation and Thickness on the Bandgap in Quasi-2D Perovskites (PEA)2(MA)n–1PbnI3n+1. *J Phys Chem C*. 2024;128(40):17091-17097. doi:10.1021/acs.jpcc.4c03540

25. Shao M, Bie T, Yang L, et al. Over 21% Efficiency Stable 2D Perovskite Solar Cells. *Adv Mater*. 2022;34(1):2107211. doi:10.1002/adma.202107211

26. Kober-Czerny M, Motti SG, Holzhey P, et al. Excellent Long-Range Charge-Carrier Mobility in 2D Perovskites. *Adv Funct Mater*. 2022;32(36):2203064. doi:10.1002/adfm.202203064

27. Blancon JC, Tsai H, Nie W, et al. Extremely efficient internal exciton dissociation through edge states in layered 2D perovskites. *Science*. 2017;355(6331):1288-1292. doi:10.1126/science.aal4211

28. Wu G, Li X, Zhou J, et al. Fine Multi-Phase Alignments in 2D Perovskite Solar Cells with Efficiency over 17% via Slow Post-Annealing. *Adv Mater*. 2019;31(42):1903889. doi:10.1002/adma.201903889

29. Choi Y, Koo D, Jeong G, et al. A vertically oriented two-dimensional Ruddlesden–Popper phase perovskite passivation layer for efficient and stable inverted perovskite solar cells. *Energy Environ Sci*. 2022;15(8):3369-3378. doi:10.1039/D2EE00759B

30. Chen AZ, Shiu M, Deng X, et al. Understanding the Formation of Vertical Orientation in Two-dimensional Metal Halide Perovskite Thin Films. *Chem Mater*. 2019;31(4):1336-1343. doi:10.1021/acs.chemmater.8b04531

31. Li X, Gu L, Feng Y, Wu C, Yang C, Song L. Vertically Oriented 2D Perovskite Layer with n = 1 Phase Enables Efficient and Stable Inverted Perovskite Solar Cells. *ACS Appl Mater Interfaces*. Published online October 22, 2024. doi:10.1021/acsami.4c14866

32. Zanetta A, Larini V, Vikram, et al. Vertically oriented low-dimensional perovskites for high-efficiency wide band gap perovskite solar cells. *Nat Commun*. 2024;15(1):9069. doi:10.1038/s41467-024-53339-6

33. Zhang X, Wu G, Yang S, et al. Vertically Oriented 2D Layered Perovskite Solar Cells with Enhanced Efficiency and Good Stability. *Small*. 2017;13(33):1700611. doi:10.1002/smll.201700611

34. Zhang H, Hou M, Xia Y, et al. Synergistic effect of anions and cations in additives for highly efficient and stable perovskite solar cells. *J Mater Chem A*. 2018;6(19):9264-9270. doi:10.1039/C8TA00308D

35. Miró-Zárate JL, Elías-Espinosa MC, Cervantes-Sodi F, Diliegros-Godines CJ. Effects of NH4SCN Additive in the FAPbI3 Perovskite Films in a Sequential Deposition Method. *Crystals*. 2023;13(5):795. doi:10.3390/cryst13050795

36. Yukta, Chavan RD, Yadav P, Satapathi S. Ammonium Thiocyanate-Passivated Quasi-Two-Dimensional Dion Jacobson Perovskite Solar Cells for Improved Efficiency and Stability. *ACS Appl Energy Mater*. 2022;5(11):13723-13734. doi:10.1021/acsaem.2c02398

37. Zhang X, Wu G, Fu W, et al. Orientation Regulation of Phenylethylammonium Cation Based 2D Perovskite Solar Cell with Efficiency Higher Than 11%. *Adv Energy Mater*. 2018;8(14):1702498. doi:10.1002/aenm.201702498

38. Gao Y, Gao X, Zhang Y, Zhang J, Jing X, Shao M. Fluorinated Low-Dimensional Ruddlesden–Popper Perovskite Solar Cells with over 17% Power Conversion Efficiency and Improved Stability. *Adv Mater*. 2019;31(37):1901673. doi:10.1002/adma.201901673

39. Fu W, Wang J, Zuo L, et al. Two-Dimensional Perovskite Solar Cells with 14.1% Power Conversion Efficiency and 0.68% External Radiative Efficiency. *ACS Energy Lett*. 2018;3(9):2086-2093. doi:10.1021/acsenergylett.8b01181

40. Duim H, Rommens AJ, Shao S, Portale G, Loi MA. Addition of Ammonium Thiocyanate Alters the Microstructure and Energetic Landscape of 2D/3D Perovskite Films. *ACS Appl Energy Mater*. 2023;6(20):10285-10293. doi:10.1021/acsaem.3c00545

41. Gao G, Dong Y, Jiang L, et al. Phase Distribution Dictates Charge Transfer and Transport Dynamics in Layered Quasi-2D Perovskite. *Nano Lett*. 2024;24(42):13356-13363. doi:10.1021/acs.nanolett.4c03814

42. Zhao X, Zheng T, Zhao W, Yu Y, Wang W, Ni Z. Photoluminescence Modulation of Ruddlesden-Popper Perovskite via Phase Distribution Regulation. *Nanomaterials*. 2023;13(3):571. doi:10.3390/nano13030571

43. Ramadan AJ, Jeong WH, Oliver RDJ, et al. The Role of the Organic Cation in Developing Efficient Green Perovskite LEDs Based on Quasi-2D Perovskite Heterostructures. *Adv Funct Mater*. 2024;34(14):2309653. doi:10.1002/adfm.202309653

44. Singh B, Saykar NG, Kumar BS, Afria D, C. K. S, Rondiya SR. Unraveling Interface-Driven and Loss Mechanism-Centric Phenomena in 3D/2D Halide Perovskites: Prospects for Optoelectronic Applications. *ACS Omega*. 2024;9(9):10000-10016. doi:10.1021/acsomega.3c08936

45. Zhang J, Wu J, Langner S, et al. Exploring the Steric Hindrance of Alkylammonium Cations in the Structural Reconfiguration of Quasi-2D Perovskite Materials Using a High-throughput Experimental Platform. *Adv Funct Mater*. 2022;32(43):2207101. doi:10.1002/adfm.202207101

46. Min H, Chang J, Tong Y, et al. Additive treatment yields high-performance lead-free perovskite light-emitting diodes. *Nat Photonics*. 2023;17(9):755-760. doi:10.1038/s41566-023-01231-y

47. Abdelsamie M, Li T, Babbe F, et al. Mechanism of Additive-Assisted Room-Temperature Processing of Metal Halide Perovskite Thin Films. *ACS Appl Mater Interfaces*. 2021;13(11):13212-13225. doi:10.1021/acsami.0c22630

48. Sun Y, Peng J, Chen Y, Yao Y, Liang Z. Triple-cation mixed-halide perovskites: towards efficient, annealing-free and air-stable solar cells enabled by Pb(SCN)2 additive. *Sci Rep*. 2017;7(1):46193. doi:10.1038/srep46193

49. K. Perera WH, Webb T, Xu Y, et al. 23.2% efficient low band gap perovskite solar cells with cyanogen management. *Energy Environ Sci*. 2025;18(1):439-453. doi:10.1039/D4EE03001J

50. Brennan MC, Forde A, Zhukovskyi M, et al. Universal Size-Dependent Stokes Shifts in Lead Halide Perovskite Nanocrystals. *J Phys Chem Lett*. 2020;11(13):4937-4944. doi:10.1021/acs.jpclett.0c01407

51. Di D, Musselman KP, Li G, et al. Size-Dependent Photon Emission from Organometal Halide Perovskite Nanocrystals Embedded in an Organic Matrix. *J Phys Chem Lett*. 2015;6(3):446-450. doi:10.1021/jz502615e

52. Haward TB, Lim VJY, Cherniukh I, Bodnarchuk MI, Kovalenko MV, Herz LM. Correlated Vibrational and Electronic Signatures of Surface Disorder in CsPbBr3 Nanocrystals. *ACS Nano*. 2025;19(46):40159-40169. doi:10.1021/acsnano.5c16045

53. Hou J, Li W, Zhang H, et al. Synthesis of 2D perovskite crystals via progressive transformation of quantum well thickness. *Nat Synth*. Published online October 23, 2023:1-11. doi:10.1038/s44160-023-00422-3

54. Zhang F, Cai B, Song J, Han B, Zhang B, Zeng H. Efficient Blue Perovskite Light-Emitting Diodes Boosted by 2D/3D Energy Cascade Channels. *Adv Funct Mater*. 2020;30(27):2001732. doi:10.1002/adfm.202001732

55. Jiang R, Tian T, Ke B, et al. Insights into the effects of oriented crystallization on the performance of quasi-two-dimensional perovskite solar cells. *Mater*. 2023;1(4):100044. doi:10.1016/j.nxmate.2023.100044

56. Pahwa N, Singh AR. “Structural, Optical and AFM based PSD surface roughness studies of cadmium telluride thin films of nanometer thickness range.” *Results Surf Interfaces*. 2025;21:100655. doi:10.1016/j.rsurfi.2025.100655

57. Karan S, Mallik B. Power spectral density analysis and photoconducting behavior in copper(II) phthalocyanine nanostructured thin films. *Phys Chem Chem Phys*. 2008;10(45):6751-6761. doi:10.1039/B809648A

58. Eftekhari L, Raoufi D, Javad Eshraghi M, Ghasemi M. Power spectral density-based fractal analyses of sputtered yttria-stabilized zirconia thin films. *Semicond Sci Technol*. 2022;37(10):105011. doi:10.1088/1361-6641/ac84fa

59. Bush KA, Rolston N, Gold-Parker A, et al. Controlling Thin-Film Stress and Wrinkling during Perovskite Film Formation. *ACS Energy Lett*. 2018;3(6):1225-1232. doi:10.1021/acsenergylett.8b00544

60. Slavney AH, Smaha RW, Smith IC, Jaffe A, Umeyama D, Karunadasa HI. Chemical Approaches to Addressing the Instability and Toxicity of Lead–Halide Perovskite Absorbers. *Inorg Chem*. 2017;56(1):46-55. doi:10.1021/acs.inorgchem.6b01336

61. G S, Mahale P, Kore BP, et al. Is CH3NH3PbI3 Polar? *J Phys Chem Lett*. 2016;7(13):2412-2419. doi:10.1021/acs.jpclett.6b00803

62. Schmitt T, Bourelle S, Tye N, et al. Control of Crystal Symmetry Breaking with Halogen-Substituted Benzylammonium in Layered Hybrid Metal-Halide Perovskites. *J Am Chem Soc*. 2020;142(11):5060-5067. doi:10.1021/jacs.9b11809

63. Steele JA, Solano E, Hardy D, et al. How to GIWAXS: Grazing Incidence Wide Angle X-Ray Scattering Applied to Metal Halide Perovskite Thin Films. *Adv Energy Mater*. 2023;13(27):2300760. doi:10.1002/aenm.202300760

64. Synergistic energy and charge transfer dynamics in LD/3D perovskite heterojunctions for optoelectronic applications. *Chem Sci*. 2026;17(23):11273-11293. doi:10.1039/d5sc05663b

65. Breiby DW, Bunk O, Andreasen JW, Lemke HT, Nielsen MM. Simulating X-ray diffraction of textured films. *J Appl Crystallogr*. 2008;41(2):262-271. doi:10.1107/S0021889808001064

66. Calabrese J, Jones NL, Harlow RL, Herron N, Thorn DL, Wang Y. Preparation and characterization of layered lead halide compounds. *J Am Chem Soc*. 1991;113(6):2328-2330. doi:10.1021/ja00006a076

67. Hao Y, Qiu Z, Zhang X, Wei Z, Yao J, Cai H. Series of 2D multilayered perovskites constructed by slicing the 3D [(CH3NH3)PbI3] with 4-fluorobenzylamine. *Inorg Chem Commun*. 2018;97:134-138. doi:10.1016/j.inoche.2018.09.030

68. Ramos-Terrón S, Argudo PG, Camacho L, et al. Unraveling the Role of MASCN in Controlling Phase Distribution for High-Efficiency Quasi-2D Stilbene-Based Perovskite Solar Cells. *ACS Energy Lett*. 2026;11(10):6872-6880. doi:10.1021/acsenergylett.6c02235

69. Metcalf I, Forte S, Zhou J, et al. Design principle for mitigating moisture-induced degradation in 2D halide perovskites. *Nat Commun*. Published online October 3, 2026. doi:10.1038/s41467-026-78054-2

70. Tang J, Tian W, Zhao C, et al. Imaging the Moisture-Induced Degradation Process of 2D Organolead Halide Perovskites. *ACS Omega*. 2022;7(12):10365-10371. doi:10.1021/acsomega.1c06989

71. Sidhik S, Metcalf I, Li W, et al. Two-dimensional perovskite templates for durable, efficient formamidinium perovskite solar cells. *Science*. 2024;384(6701):1227-1235. doi:10.1126/science.abq6993

72. Wang Y, Shi Z, Wang Y, et al. Highly Oriented FAPbI3 via 2D Ruddlesden Popper Perovskite Template Growth. *Adv Energy Mater*. 2024;14(31):2401721. doi:10.1002/aenm.202401721

73. Luo C, Zheng G, Gao F, et al. Facet orientation tailoring via 2D-seed- induced growth enables highly efficient and stable perovskite solar cells. *Joule*. 2022;6(1):240-257. doi:10.1016/j.joule.2021.12.006

74. Wang J, Luo S, Lin Y, et al. Templated growth of oriented layered hybrid perovskites on 3D-like perovskites. *Nat Commun*. 2020;11(1):582. doi:10.1038/s41467-019-13856-1

75. Wu Y, Yang S, Li E, et al. Low-dimensional templates and delayed crystallization for high-quality tin-based perovskite films and high-performance transistors. *Nat Commun*. 2025;16(1):9505. doi:10.1038/s41467-025-64560-2

76. Hou J, Li W, Zhang H, et al. Deterministic synthesis of phase pure 2D perovskites via progressive transformation of layer thickness. *arXiv*. Preprint posted online December 2, 2022:arXiv:2212.00989. doi:10.48550/arXiv.2212.00989

77. Hu X, Zhao C, Liu Y, et al. Blade-Coated Perovskite-Organic Tandem Solar Cells in Ambient Conditions. *Adv Funct Mater*. 2026;36(8):e12093. doi:10.1002/adfm.202512093

78. Bi L, Wang J, Fu Q, Jen AKY. Crystallization Modulation for Stable Wide-Bandgap Perovskite Solar Cells and Modules. *Adv Mater*. 2026;38(55):e74388. doi:10.1002/adma.74388

79. Li Z, Li X, Wang M, et al. Enhanced Photovoltaic Performance via a Bifunctional Additive in Tin-Based Perovskite Solar Cells. *ACS Appl Energy Mater*. 2022;5(1):108-115. doi:10.1021/acsaem.1c02488

80. Subha TD, Prabu RT, Parasuraman S, Kumar A. Role of Urbach energy in controlling voltage output of solar cells. *Opt Quantum Electron*. 2023;55(9):794. doi:10.1007/s11082-023-05067-2

81. Caprioglio P, Stolterfoht M, Wolff CM, et al. On the Relation between the Open-Circuit Voltage and Quasi-Fermi Level Splitting in Efficient Perovskite Solar Cells. *Adv Energy Mater*. 2019;9(33):1901631. doi:10.1002/aenm.201901631

82. Stolterfoht M, Caprioglio P, Wolff CM, et al. The impact of energy alignment and interfacial recombination on the internal and external open-circuit voltage of perovskite solar cells. *Energy Environ Sci*. 2019;12(9):2778-2788. doi:10.1039/C9EE02020A

83. Wolff CM, Caprioglio P, Stolterfoht M, Neher D. Nonradiative Recombination in Perovskite Solar Cells: The Role of Interfaces. *Adv Mater*. 2019;31(52):1902762. doi:10.1002/adma.201902762

## Acknowledgements

The authors acknowledge funding from the Engineering and Physical Sciences Research Council (EPSRC) (EP/X039285/1), (EP/V027131/1) (E.J.C.), (EP/R00661X/1, EP/S019367/1, EP/P02470X/1 and EP/P025285/1) (Henry Royce Institute for Advanced Materials, A.L.B.P.). The authors are grateful for funding from the EPSRC to purchase (EP/M028437/1) and upgrade (EP/V034804/1) the Xeuss 2.0 SAXS/WAXS laboratory beamline at the Soft Matter Analytical Laboratory, University of Sheffield. A. L. B. P. acknowledges the UKRI Future Leaders Fellowship MR/W00738X/1. S.T. and M.A. acknowledge the Grantham Centre for Sustainable Futures for funding. A.D.M. acknowledges funding from the National Science Foundation Science and Technology Center (STC) for Integration of Modern Optoelectronic Materials on Demand (IMOD) under Grant No. DMR-2019444. I.M. acknowledges financial support from the Hertz Foundation and the National Science Foundation Graduate Research Fellowship Program. J.E. acknowledges financial support from the Institut Universitaire de France.

## Author Contributions

Sophie Tucker: conceptualization, methodology, investigation, formal analysis, writing – original draft, review and editing.
Isaac Metcalf: investigation, formal analysis, methodology, writing- review and editing
Niranjana Raj: investigation, writing- review and editing
Jessica Persaud: investigation, formal analysis, writing- review and editing
Faiz Mandani: investigation, formal analysis, writing- review and editing
Rachel Kilbride: investigation, formal analysis, methodology, writing- review and editing
Elena J. Cassella: investigation, methodology, writing- review and editing
Max Allen: investigation, writing- review and editing
Charlie Nicholls: investigation, writing- review and editing
David G. Lidzey: funding acquisition, supervision, writing- review and editing
Alice L. B. Pyne: funding acquisition, supervision, writing- review and editing
Robert D. J. Oliver: conceptualization, supervision, writing- review and editing

Silvia Motti: funding acquisition, supervision, writing- review and editing
Jacky Even: conceptualization, supervision, writing- review and editing
Aditya Mohite: conceptualization, funding acquisition, supervision, writing- review and editing
Alexandra J. Ramadan: conceptualization, funding acquisition, methodology, supervision, writing – original draft, review and editing.

## Conflict of Interest

D. G. L. is co-chairman of the materials science company Ossila Ltd. which retails materials and equipment for photovoltaic research, including perovskite photovoltaics.

## Data Availability Statement

All raw data relating to this work is available at 10.15131/shef.data.34830888.